\documentclass[aps,prd,twocolumn,showpacs,preprintnumbers,nofootinbib]{revtex4}

\usepackage{graphicx}% Include figure files
\usepackage[FIGTOPCAP]{subfigure}% allow subfloats
\usepackage{dcolumn}% Align table columns on decimal point
\usepackage{bm}% bold math
\usepackage[normalem]{ulem}

\usepackage{amsmath}
\usepackage{epstopdf}
\usepackage{amsmath,amssymb}

\newcommand{\beq}{\begin{eqnarray}}
\newcommand{\eeq}{\end{eqnarray}}

\begin{document}

\title{$\theta$ dependence in $SU(3)$ Yang-Mills theory from 
analytic continuation}

\author{Claudio Bonati}
\email{claudio.bonati@df.unipi.it}
\affiliation{Dipartimento di Fisica dell'Universit\`a di Pisa, Largo Pontecorvo 3, I-56127 Pisa, Italy}
\affiliation{INFN - Sezione di Pisa, Largo Pontecorvo 3, I-56127 Pisa, Italy}

\author{Massimo D'Elia}
\email{massimo.delia@unipi.it}
\affiliation{Dipartimento di Fisica dell'Universit\`a di Pisa, Largo Pontecorvo 3, I-56127 Pisa, Italy}
\affiliation{INFN - Sezione di Pisa, Largo Pontecorvo 3, I-56127 Pisa, Italy}

\author{Aurora Scapellato}
\email{scapellato.aurora@ucy.ac.cy}
\altaffiliation[Present Address: ]{Department of Physics, University of Cyprus, P.O. Box 20537, 1678 Nicosia, Cyprus and
Department of Physics, Bergische Universit\"{a}t Wuppertal Gaussstr. 20, 42119 Wuppertal, Germany.}
\affiliation{Dipartimento di Fisica dell'Universit\`a di Pisa, Largo Pontecorvo 3, I-56127 Pisa, Italy}
\affiliation{INFN - Sezione di Pisa, Largo Pontecorvo 3, I-56127 Pisa, Italy}

\date{\today}% It is always \today, today,

\begin{abstract}
We investigate the topological properties of the $SU(3)$ pure gauge theory by
performing numerical simulations at imaginary values of the $\theta$ parameter.
By monitoring the dependence of various cumulants of the topological charge
distribution on the imaginary part of $\theta$ and exploiting analytic
continuation, we determine the free energy density up to the sixth order 
in $\theta$, $f(\theta,T) = f(0,T) + {1\over 2} \chi(T) \theta^2 (1 + b_2(T) \theta^2 +
b_4(T) \theta^4 + O(\theta^6))$. That permits us to achieve determinations with
improved accuracy, in particular for the higher order terms, with control over
the continuum and the infinite volume extrapolations.  We obtain
$b_2=-0.0216(15)$ and $|b_4|\lesssim 4\times 10^{-4}$.
\end{abstract}


\maketitle

\section{Introduction}\label{sec:intro}

The non-trivial consequences related to the possible presence of a topological
parameter $\theta$ are among the most interesting properties of non-abelian
gauge theories. That enters the Euclidean Yang-Mills Lagrangian as follows
\begin{equation}
{\cal L}_\theta  = \frac{1}{4} F_{\mu\nu}^a(x)F_{\mu\nu}^a(x)
- i \theta \frac{g^2}{64\pi^2} \epsilon_{\mu\nu\rho\sigma}
F_{\mu\nu}^a(x) F_{\rho\sigma}^a(x),
\label{lagrangian}
\end{equation}
where
\begin{equation}
q(x)=\frac{g^2}{64\pi^2} 
\epsilon_{\mu\nu\rho\sigma} F_{\mu\nu}^a(x) F_{\rho\sigma}^a(x)
\label{topchden}
\end{equation}
is the topological charge density; $\theta$ is a superselected parameter,
characterizing the vacuum as well as the other physical states of the theory.
The topological charge density is odd under $CP$ symmetry, hence a non-zero
$\theta$ value would break such symmetry explicitly.

Experimental upper bounds on $\theta$ are quite stringent ($|\theta| \lesssim
10^{-10}$), nevertheless, the dependence of Quantum ChromoDynamics (QCD) and of
$SU(N)$ gauge theories on $\theta$ enters various aspects of hadron
phenomenology, a paradigmatic example being the solution of the $U_A(1)$
problem \cite{'tHooft:1976up, Witten:1979vv, Veneziano:1979ec}.  By CP symmetry
at $\theta =0$, the free energy density $f$ of the theory is an even function
of $\theta$ which can be parametrized as follows
\begin{eqnarray}\label{eq:s}
f(\theta,T) = f(0,T) + {1\over 2} \chi(T)
\theta^2 s(\theta,T)
\end{eqnarray}
where  $\chi(T)$ is the topological susceptibility, while  $s(\theta,T)$ is a
dimensionless even function of $\theta$, normalized as $s(0,T)=1$, which,
assuming analyticity around $\theta=0$, can be expanded as follows
\begin{eqnarray}\label{eq:s_bis}
s(\theta,T) = 1 + b_2(T) \theta^2 + b_4(T) \theta^4 + \cdots \, .
\label{stheta}
\end{eqnarray}

The dependence on $\theta$, being related to the topological properties of the
theory, is of inherent non-perturbative nature. Therefore analytic predictions
are restricted to particular regimes or effective approximation schemes
\cite{Gross:1980br, schschrev, Zhitnitsky:2008ha, Unsal:2012zj, Poppitz:2012nz,
Anber:2013sga, Bigazzi:2015bna}.  For instance,  at asymptotically high
temperatures the Dilute Instanton Gas Approximation is expected to hold, which
predicts $f(\theta,T)-f(0,T)\simeq \chi(T)(1-\cos\theta)$, while regarding the
low-temperature phase and the $\theta$-dependence of the ground state energy,
large-$N$ arguments~\cite{Witten-98,Witten-80} predict the various non-linear terms in $\theta^2$ to be
suppressed by increasing powers of $1/N$, in particular $b_{2n} \propto
1/N^{2n}$ (see e.g. \cite{Vicari:2008jw} for a general review of the subject).

Lattice QCD represents a valid non-perturbative approach, which is based on
first principles only, however the complex nature of the Euclidean action in
Eq.~\eqref{lagrangian} hinders direct Monte-Carlo simulations at non-zero
$\theta$.  Some possible, partial solutions to this problem are similar to the
ones adopted for QCD at finite baryon chemical potential $\mu_B$, where the
fermion determinant becomes complex.  In particular, assuming analyticity
around $\theta = 0$, one can either obtain the free energy density in terms of
its Taylor expansion coefficients computed at $\theta = 0$ (see
Refs.~\cite{taylormu1, taylormu2, taylormu3} for analogous strategies at $\mu_B
\neq 0$), or perform numerical simulations at imaginary values of
$\theta$~\cite{azcoiti,alles_1,alles_2,aoki_1,pavim,dene} (or $\mu_B$ \cite{immu1,
immu2, immu3, immu4, mdfs1, tadena}) and then exploit analytic continuation.

The first strategy has been traditionally adopted for the study of $\theta$
dependence.  The topological susceptibility $\chi(T)$ and the coefficients
$b_{2n}$ are proportional to the coefficients of the Taylor expansion in
$\theta$ and can be directly computed in terms of the cumulants of the
probability distribution at $\theta=0$ of the global topological charge $Q$.
For instance the first few terms are given by
\begin{equation}\label{eq:chi}
\chi = \frac{\langle Q^2 \rangle_{\theta=0}}{\mathcal{V}}\end{equation}
where $\mathcal{V}$ is the four-dimensional volume, 
\begin{equation}\label{eq:b2}
b_2=-\frac{\langle Q^4\rangle_{\theta=0}-
         3\langle Q^2\rangle^2_{\theta=0}}{12\langle Q^2\rangle_{\theta=0}} \, ,
\end{equation}
and
\begin{equation}\label{b_4}
b_4=\frac{\left[ \langle Q^6\rangle-15\langle Q^2\rangle \langle Q^4\rangle 
+30\langle Q^2\rangle ^3\right]_{\theta=0}}{360 \langle Q^2\rangle_{\theta=0}} \ .
\end{equation}

As we will discuss in more detail in the following, a drawback of this approach
is that the coefficients of the non-quadratic terms, $b_{2n}$, are determined
as corrections to a purely gaussian behavior for the probability distribution
of the topological charge at $\theta = 0$. By the central limit theorem, such
corrections are less and less visible as the total four-dimensional volume 
increases, so that a significant increase in statistics is needed, in order
to keep a constant statistical error, as one tries to increase $\mathcal{V}$
in order to keep finite size effects under control.

This problem can be avoided with the approach based on analytic continuation
from an imaginary $\theta$. In this case, one typically determines a derivative
of the free energy as a function of $\theta = -i \theta_I$: for instance one
has
\begin{equation}\label{eq:fprime}
\frac{\langle Q \rangle_{\theta_I}}{\mathcal{V}} 
= \chi\, \theta_I (1 - 2 b_2 \theta_I^2) + O(\theta_I^5)
\end{equation}
and $b_2$ is obtained as a deviation from linearity of the response of the
system to the external source $\theta_I$, which is determined with no loss in
accuracy as the system size is increased.  A drawback in  this case is
represented by a finite renormalization appearing when the external source
$\theta_I$ is added to the discretized theory; nevertheless, the method of
analytic continuation turns out to be the most suitable to a high precision
study of the coefficients $b_{2n}$, which must keep both finite size and
discretization errors under control.

In this study we investigate the $\theta$-dependence of $SU(3)$ pure gauge
theory using the analytic continuation approach.  We will explore variants of
the original strategy presented in Ref.~\cite{pavim}, in particular we will
perform a simultaneous fit to various derivatives of the free energy density as
a function of $\theta_I$, in analogy with a similar approach explored in
lattice QCD at imaginary chemical potential~\cite{tadena}.  That will permit us
to maximize the information obtained from our numerical simulations and, at the
same time, to determine the finite renormalization constant as a parameter of
the global fit. That will result in an increased precision, which will give us
the opportunity to determine a continuum extrapolated value of $b_2$ with an
uncertainty at the level of a few percent.

\section{Numerical method}

%\subsection{Imaginary theta and analytic continuation}
\label{sec:imtheta}

The free energy density of Yang-Mills theory in the presence of a $\theta$ term 
is given by
\begin{equation}\label{eq:f}
e^{-\mathcal{V} f(\theta,T)}=\int [dA]\, e^{- (S_{YM}^{E}[A]-i\theta Q[A])} \ ,
\end{equation}
where $S_{YM}^{E}$ is the standard euclidean action of Yang-Mills theory,
$Q=\int q(x)\mathrm{d} x$ is the topological charge and periodic boundary
conditions are assumed.
As discussed in the
introduction it is not possible to use directly this form in a Monte Carlo
simulation, since the action is not real for $\theta\neq 0$.  To overcome this
problem we will study the theory for imaginary values of $\theta$, where
standard importance sampling methods can be applied, using the analytic
continuation of Eqs.\eqref{eq:s}, \eqref{eq:s_bis} to extract the values of the
coefficients.

In practice, one defines $\theta\equiv -i \theta_I$,
where $\theta_I$ is a real parameter, and studies the $\theta_I$ dependence of
\begin{equation}\label{eq:F}
\begin{aligned}
& \tilde f (\theta_I,T)\equiv f(-i\theta_I, T)-f(0,T)=\\
& =-\frac{1}{2}\chi \theta_I^2 (1-b_2\theta_I^2+b_4\theta_I^4+\ldots )\ .
\end{aligned}
\end{equation}
From Eq.~\eqref{eq:f} it follows that the derivatives of $\tilde f$ can be 
written as
\begin{equation}\label{eq:derivate}
\frac{\partial^k \tilde f 
(\theta_I)}{\partial\theta_I ^k}=-\frac{1}{\mathcal{V}}
\langle Q^k\rangle _{c,\theta_I}\ ,
\end{equation}
where $\langle Q^k\rangle _{c,\theta_I}$ are the cumulants of the topological
charge distribution at fixed $\theta_I$. Since for $\theta_I\neq 0$ the CP
symmetry is explicitly broken, also the odd cumulants are generically
non-vanishing.  Using Eq.~\eqref{eq:derivate} together with the expansion in
Eq.~\eqref{eq:F} we obtain an infinite chain of equations relating the
cumulants of the topological charge to the coefficients $\chi$ and $b_{2n}$.
The first few of these equations are  the following
\begin{equation} \label{eq:cum_theta}
\begin{split}
\frac{\langle Q\rangle_{c,\theta_I}}{\mathcal{V}}&=
\chi\theta_I(1-2b_2\theta_I^2+3b_4\theta_I^4+\ldots ),\\
\frac{\langle Q^2\rangle_{c,\theta_I}}{\mathcal{V}}&=
\chi(1-6b_2\theta_I^2+15b_4\theta_I^4+\ldots ),\\
\frac{\langle Q^3\rangle_{c,\theta_I}}{\mathcal{V}}&=
\chi (-12b_2\theta_I+60b_4\theta_I^3+\ldots ),\\
\frac{\langle Q^4\rangle_{c,\theta_I}}{\mathcal{V}}&=
\chi (-12b_2+180b_4\theta_I ^2+\ldots ).
\end{split}
\end{equation}
The general idea of the analytic continuation method is to perform Monte Carlo
simulations at several values of $\theta_I$, to compute some of the cumulants
$\langle Q^k\rangle_{c,\theta_I}$ for each $\theta_I$ and to extract the
coefficients $\chi$, $b_{2n}$ making use Eq.~\eqref{eq:cum_theta}.

It is clear that in this procedure one can adopt different strategies. Each of
the relations in Eq.~\eqref{eq:cum_theta} is itself sufficient to extract
all the parameters, at least from a given order on: for example $\chi$ can be estimated by looking at the
leading linear dependence of $\langle Q\rangle_{c,\theta_I}$ on $\theta_I$ or
by looking at the small $\theta_I$ value of $\langle Q^2\rangle_{c, \theta_I}$.
While all these methods are equally valid from
the theoretical point of view, they are not equivalent in computational
efficiency, since 
at fixed statistics lower cumulants are determined with better
accuracy (see Sec.~\ref{comparison} for a detailed discussion on this point).
On the other hand the computation of higher cumulants does not require new
simulations or even new measurements, so that if we use \emph{also} higher
momenta we increase the information available at no additional cost. What
appears to be the optimal strategy is to extract $\chi$ and $b_{2n}$ from a
combined fit of all the relations in Eq.~\eqref{eq:cum_theta}. Since
higher momenta get more and more noisy, it is natural to expect that, at some
point, adding to the global fit still higher cumulants will not result in an
increased precision, however it is not a priori clear when this will happen; in
the following analysis we will use up to the fourth cumulant.  Obviously, in
order to correctly assess the statistical uncertainties of the extracted
parameters, correlations among the cumulants have to be taken into account.

\subsection{Lattice implementation and topological charge definition}\label{sec:top_lat}

Various methods exist to determine the topological content of lattice
configurations, either based on fermionic or gluonic definitions, which have
proven to provide consistent results as the continuum limit is approached (see
Ref.~\cite{Vicari:2008jw} for a review).  Gluonic definitions are typically
affected by renormalizations related to fluctuations at the ultraviolet (UV)
cutoff scale, which can be cured by a proper smoothing of gauge configurations.
Cooling \cite{cooling1, cooling2, cooling3, cooling4, cooling5}  is a standard
procedure adopted to do that; recently the gradient flow \cite{Luscher:2009eq,
Luscher:2010iy} has been introduced, which has been shown to provide
results equivalent to cooling, at least regarding
topology~\cite{Bonati:2014tqa, Cichy:2014qta, Namekawa:2015wua, Alexandrou:2015yba}.

However, for the purpose of inserting a $\theta$-term in the action, the use of
a fermionic or of a smoothed gluonic definition of $Q$ is impractical.  The
lattice partition function in the presence of an imaginary $\theta$-term reads
\begin{equation}
Z_L(T,\theta_L) = 
\int [dU]\ e^{ -S_L [U] + \theta_{L} Q_L[U]} \, ,
\label{partfunlat}
\end{equation}
where $U$ stands for the gauge link variables, $U_\mu (n)$, $S_L$ is the
lattice pure gauge action and $Q_L = \sum_x q_L(x)$ is the discretized
topological charge.  Standard efficient algorithms like heat-bath, which are
available for pure gauge theories, are applicable in this case only if a
particularly simple discretization of the topological charge density $q_L(x)$
is chosen, while other choices would lead to a significant computational
overhead. In our case, we choose the standard Wilson plaquette action for $S_L$
and the simplest discretization of $q(x)$ with definite parity
\cite{DiVecchia:1981qi, DiVecchia:1981hh}:
\begin{equation}\label{eq:qlattice}
q_L(x) = \frac{-1}{2^9 \pi^2} 
\sum_{\mu\nu\rho\sigma = \pm 1}^{\pm 4} 
{\tilde{\epsilon}}_{\mu\nu\rho\sigma} \hbox{Tr} \left( 
\Pi_{\mu\nu}(x) \Pi_{\rho\sigma}(x) \right) \; ,
\end{equation}
where $\Pi_{\mu\nu}$ is the plaquette operator,
${\tilde{\epsilon}}_{\mu\nu\rho\sigma} = {{\epsilon}}_{\mu\nu\rho\sigma} $ for
positive directions, while ${\tilde{\epsilon}}_{\mu\nu\rho\sigma} =
- {\tilde{\epsilon}}_{(-\mu)\nu\rho\sigma}$ and antisymmetry fix its value in
the generic case.  With this choice, gauge links still appear linearly in the
Boltzmann weight, so that standard heat-bath and over-relaxation algorithms can
be implemented, with a computational cost that is about a factor 2.5 higher
than at $\theta_L = 0$.

A drawback of this choice is that $q_L(x)$ takes a finite multiplicative
renormalization with respect to the continuum density
$q(x)$~\cite{Teper:1989ig, Campostrini:1989dh, DiGiacomo:1989id,
DiGiacomo:1991ba}
\begin{equation}
q_L(x) {\buildrel {a \rightarrow 0} \over \sim} a^4 Z(a)q(x) \; ,
\end{equation}
where $a$ is the lattice spacing and $\lim_{a \to 0} Z(a) = 1$. 
Hence, the lattice parameter $\theta_L$ is related to 
the imaginary part of $\theta$ by
$\theta_I = Z\, \theta_L$: in order to know $\theta_I$, one
has to determine the value $Z$.

As for the operator used to determine the cumulants of the topological charge
distribution, in order to avoid the appearance of further renormalization
constants, we adopt a smoothed definition which is denoted simply as $Q$ in the
following.  In particular, we adopt the topological charge density in
Eq.~\eqref{eq:qlattice}, measured after $15$ cooling steps.  Actually, in
this case two possible prescriptions can be taken. One can simply define $Q =
Q_L^{\rm smooth}$: in this case, at finite lattice spacing, $Q$ does not take
exactly integer values, although its distribution is characterized by sharp
peaks located at approximately integer values.  In alternative, one can define
$Q$ as follows~\cite{DelDebbio:2002xa}:
\begin{equation}\label{eq:Q_round}
Q=\mathrm{round}\left(\alpha\, Q_L^{\rm smooth} \right)\ ,
\end{equation} 
where $\mathrm{round}(x)$ denotes the integer closest to $x$ and the rescaling
factor $\alpha$ is determined in such a way to minimize
\begin{equation}
\left\langle \left( \alpha\, Q_L^{\rm smooth} - \mathrm{round}
\left[\alpha\, Q_L^{\rm smooth} \right]\right)^2\right\rangle \ ,
\end{equation}
so that the sharp peaks in the distribution of $\alpha Q_L^{\rm smooth}$ will
be located exactly onto integer values. The two definitions are expected to
coincide as the continuum limit is approached.  In the following we will refer
to the latter as the {\em rounded} topological charge and to the former as the
{\em non-rounded} one. The two definitions have been used alternately in
various previous studies in the literature.  In this study we consider both of
them and show that, while results at finite lattice spacing differ, continuum
extrapolated results coincide, as expected.

In order to make use of Eqs.~\eqref{eq:cum_theta} to obtain the free
energy parameters from the $\theta$-dependence of the cumulants of $Q$, one
needs to know the values of $\theta_I$ used for each numerical simulation. That
in turn requires a determination of the renormalization constant $Z$. Various
methods exist to do that, for instance by using the relation~\cite{pavim}
\begin{equation}\label{eq:Zdet}
Z \equiv \frac{\langle Q Q_L \rangle_{\theta=0}}{\langle Q^2 \rangle_{\theta=0}}
\end{equation}
or other similar approaches, like heating techniques~\cite{DiGiacomo:1991ba}.
However, a drawback of this approach is that the statistical uncertainty on $Z$
propagates to $\theta_I$, so that a non-trivial fit with errors on both
dependent and independent variables is required. For that reason in this study
we investigate and adopt a different approach.

We rewrite a lattice version of Eqs.~\eqref{eq:cum_theta} in which
the lattice parameter $\theta_L$ appears explicitly:
\begin{equation}\label{eq:cum_theta_L}
\begin{split}
\frac{\langle Q \rangle}{\mathcal{V}}&=
\chi Z \theta_L (1 - 2 b_2 Z^2 \theta_L^2 + 3 b_4 Z^4 \theta_L^4 + \dots)\, , \\
\frac{\langle Q^2 \rangle_c}{\mathcal{V}}&=  
\chi (1 - 6 b_2 Z^2 \theta_L^2 + 15 b_4 Z^4 \theta_L^4 + \dots)\, , \\
\frac{\langle Q^3 \rangle_c}{\mathcal{V}}&=  
\chi (- 12 b_2 Z \theta_L + 60 b_4 Z^3 \theta_L^3 + \dots)\, , \\ 
\frac{\langle Q^4 \rangle_c}{\mathcal{V}}&=
\chi (- 12 b_2 + 180 b_4 Z^2 \theta_L^2 + \dots)\, .
\end{split}
\end{equation}
Our proposal is to extract the renormalization constant from the fitting
procedure itself, treating $\theta_L$ as the independent variable and $Z$ as a
fit parameter.  Notice however that, in order to do that, at least two
cumulants must be fitted simultaneously, otherwise $Z$ remains entangled as an
arbitrary multiplicative renormalization of the other fit parameters.  On the
other hand, fitting the dependence of more cumulants at the same time coincides
with our proposed strategy to extract the best possible information from our
Monte-Carlo simulations.  As we will show in the next section, the payoff of
this strategy is not negligible.

\section{Numerical results}

\begin{table}
\begin{tabular}{|l|l|l|}
\hline $\beta$ & $r_0/a$ & $a$ [fm] \\ \hline
5.95 & 4.898(12) & 0.1021(25)  \\ \hline
6.07 & 6.033(17) & 0.08288(23) \\ \hline
6.20 & 7.380(26) & 0.06775(24) \\ \hline
6.40 & 9.74(5)   & 0.05133(26) \\ \hline
\end{tabular}
\caption{Physical scale determination at the bare couplings used in this work,
from Ref.~\cite{Guagnelli:1998ud}. $r_0$ is the Sommer parameter
\cite{Sommer:1993ce} and in the last column the lattice spacing is obtained by
using $r_0\simeq 0.5\,\mathrm{fm}$.}\label{tab:a}
\end{table}

\begin{table}
\begin{tabular}{|l|l|l|}
\hline\rule{0mm}{3mm}$\beta$ & $N_t\times N_s^3$ & $\theta_L$ \\ \hline
\rule{0mm}{3mm}5.95    & $16^4$    & 0,2,4,6,8    \\ \hline
\rule{0mm}{3mm}6.07    & $22^4$    & 0,2,4,6,8,10 \\ \hline
\rule{0mm}{3mm}6.20    & $16^4$, $18^4$, $20^4$, $22^4$    & 0,2,4,6,8,10,12 \\ \hline
\rule{0mm}{3mm}6.40    & $30^4$    & 0,2,4,6,8,10,12,14 \\ \hline \hline
\rule{0mm}{3mm}6.173   & $10\times 40^3$ & 0,2,4,6,8,10,12 \\ \hline
\end{tabular}
\caption{Bare couplings, lattice sizes and fit ranges adopted.}\label{tab:theta}
\end{table}

We have performed simulations at four different lattice spacings 
on hypercubical lattices;
the values of the bare parameters adopted are reported in
Tabs.~\ref{tab:a}-\ref{tab:theta}.  The Monte Carlo updates were performed
using a mixture of standard heatbath~\cite{Creutz1980, KP} and 
overrelaxation~\cite{Creutz1987} algorithms, implemented \emph{\`a la} Cabibbo-Marinari~\cite{CM}, in the ratio of 1 to 5. 
The topological charge was measured every
$10$ updates and from 5 to 25 cooling steps were used: 
data that will be presented in the following refer to
the case of 15 cooling steps and we checked that
results are stable, well within errors, if a different choice is made.

\begin{figure}[t!]
\includegraphics[width=0.92\columnwidth, clip]{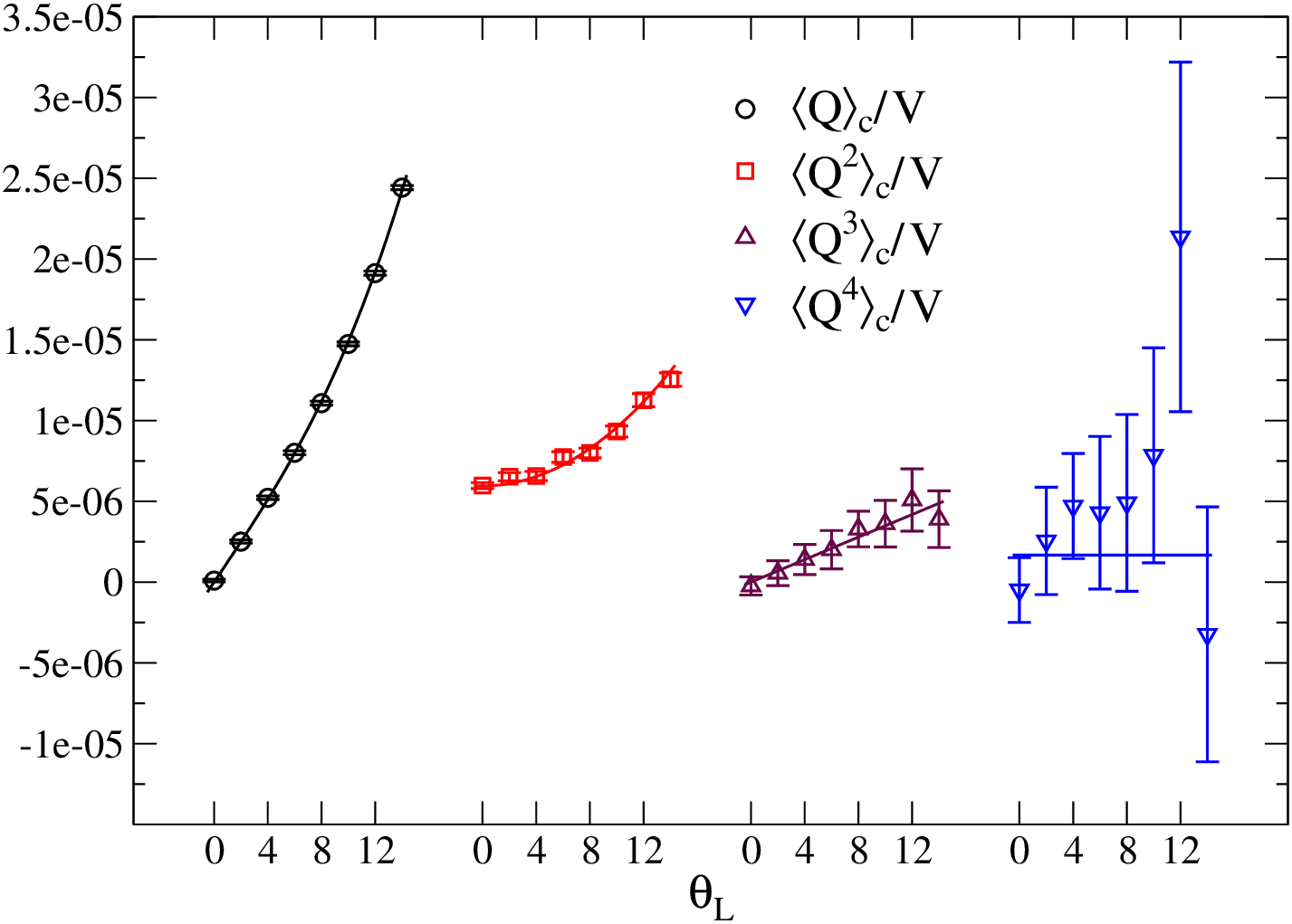}
\caption{An example of the global fit procedure: data refer to the $30^4$ lattice at coupling $\beta=6.40$, 
continuous lines are the result of a combined fit of the first four cumulants.}
\label{fig:globalfit}
\end{figure}

Unless otherwise explicitly stated, we present data
which refer to the case of a common fit to the first
four cumulants of the topological charge, see 
Eq.~(\ref{eq:cum_theta_L}). An example of such a fit is 
reported in Fig.~\ref{fig:globalfit}. 
Since cumulants of different order are computed on the same
samples of gauge configurations, in order to take
into account the possible correlations among them
we have used a bootstrap procedure, and checked that
a correlated fit exploiting the full covariance matrix
leads to consistent results.

In the following we 
are going to illustrate our estimation of the various possible
systematic uncertainties that may affect our results. First, we will analyze
those related to analytic continuation itself, by considering the stability
of our fits as either the fit range or the number of fitted terms
in Eq.~(\ref{eq:cum_theta_L}) (i.e. the truncation of the Taylor series)
is changed; then we will turn to the analysis of the infinite volume
and continuum limit extrapolations.
That will permit us to provide final estimates of the free energy 
coefficients entering Eqs.~(\ref{eq:s}) and (\ref{stheta}), with 
a reliable determination of both statistical and systematic uncertainties.
In the last part of our analysis we will try to understand which aspects
of our strategy lead to a more significant improvement with respect
to other methods used in the past literature.\\

\begin{figure}[t!]
\includegraphics[width=0.92\columnwidth, clip]{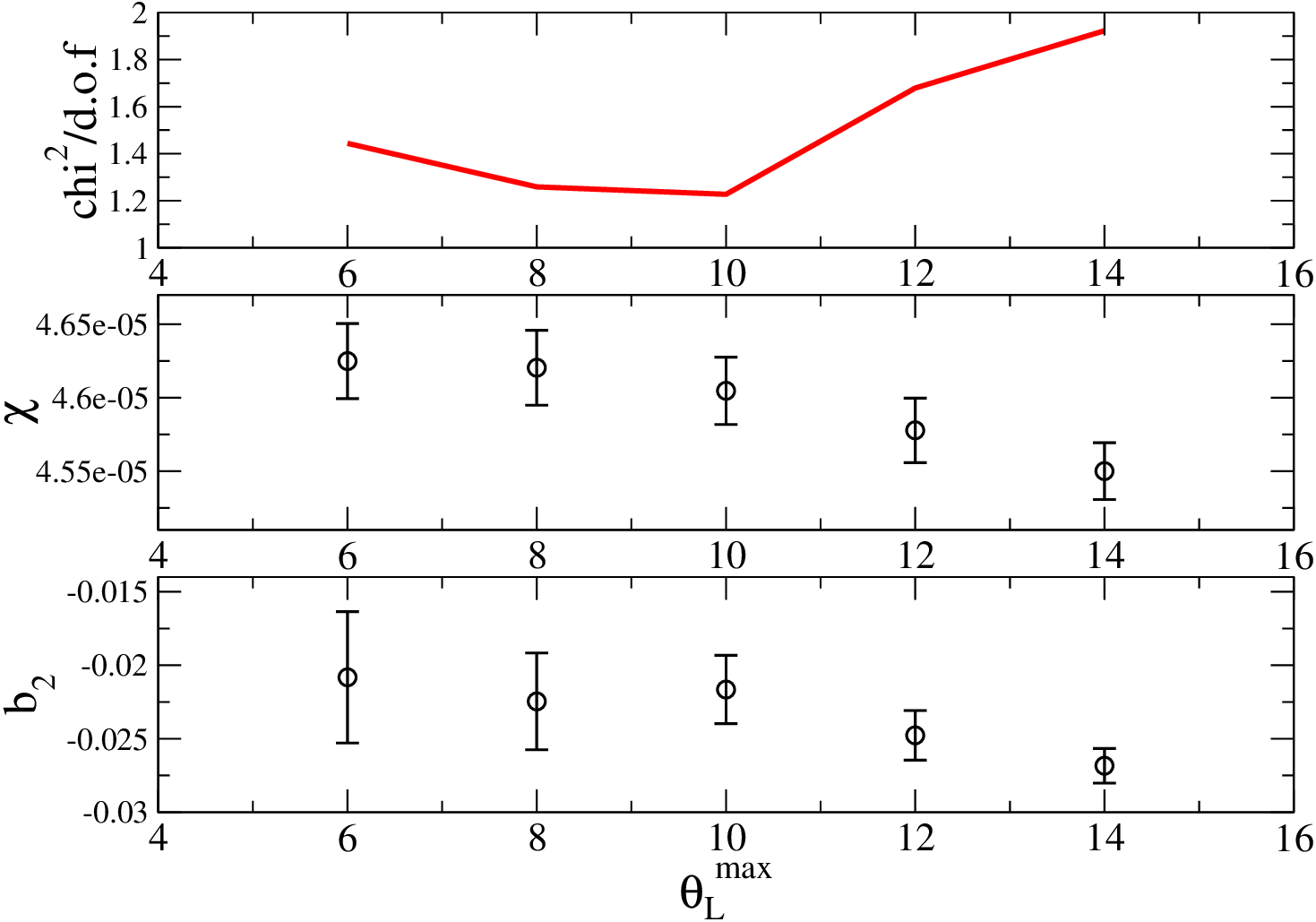}
\caption{An example of the variation of the fit quality and parameters 
with the fit range. Data refer to the $22^4$ lattice at coupling $\beta=6.07$.}
\label{fig:fitrange}
\end{figure}

In Fig.~\ref{fig:fitrange} we report an example of how
the fitted coefficients and the quality of the fit change
as a function of the range of imaginary values of 
$\theta$ included in the fit, in particular of the maximum 
value $\theta_L^{\rm max}$. The final range from which we
extract our determination is chosen so as to have 
a reasonable value of the $\chi^2/{\rm d.o.f.}$ test
and stability, within errors, of the fitted parameters
as the range is reduced further: this is important
to ensure the reliability of analytic continuation, since the expressions
in Eq.~(\ref{eq:cum_theta_L}) are expected to hold, in principle, 
in a limited range around $\theta_L = 0$.
A full account of 
the ranges used in the various cases is reported
in Tab.~\ref{tab:theta}.

We notice that $\theta_L^{\rm max}$ grows as one approaches
the continuum limit. That can be interpreted by observing that,
since growing values of $\theta_L$ induce growing values of
the net non-zero topological charge density, 
one may expect saturation effects to be present 
for large enough values of $\theta_L$. Such effects are expected
to appear earlier on lattices with less resolution, i.e. with 
larger values of the lattice spacing.

A source of systematic error that is generically present in the analytic
continuation approach is the one related to the choice 
of the fitting function, in particular to the truncation of the series to be
fitted. In the present work, the estimated value of $b_4$ turns out to be
compatible with zero for 
all the data sets, so it appears
reasonable to fix the value of $b_4$ to zero and fit just $Z$, $\chi$ and
$b_2$. To check that this does not introduce any significant 
systematic error we
verified that fits with fixed $b_4=0$ well describe data and give results
compatible with the ones obtained by fitting also $b_4$.  

\begin{figure}[t!]
\includegraphics[width=0.92\columnwidth, clip]{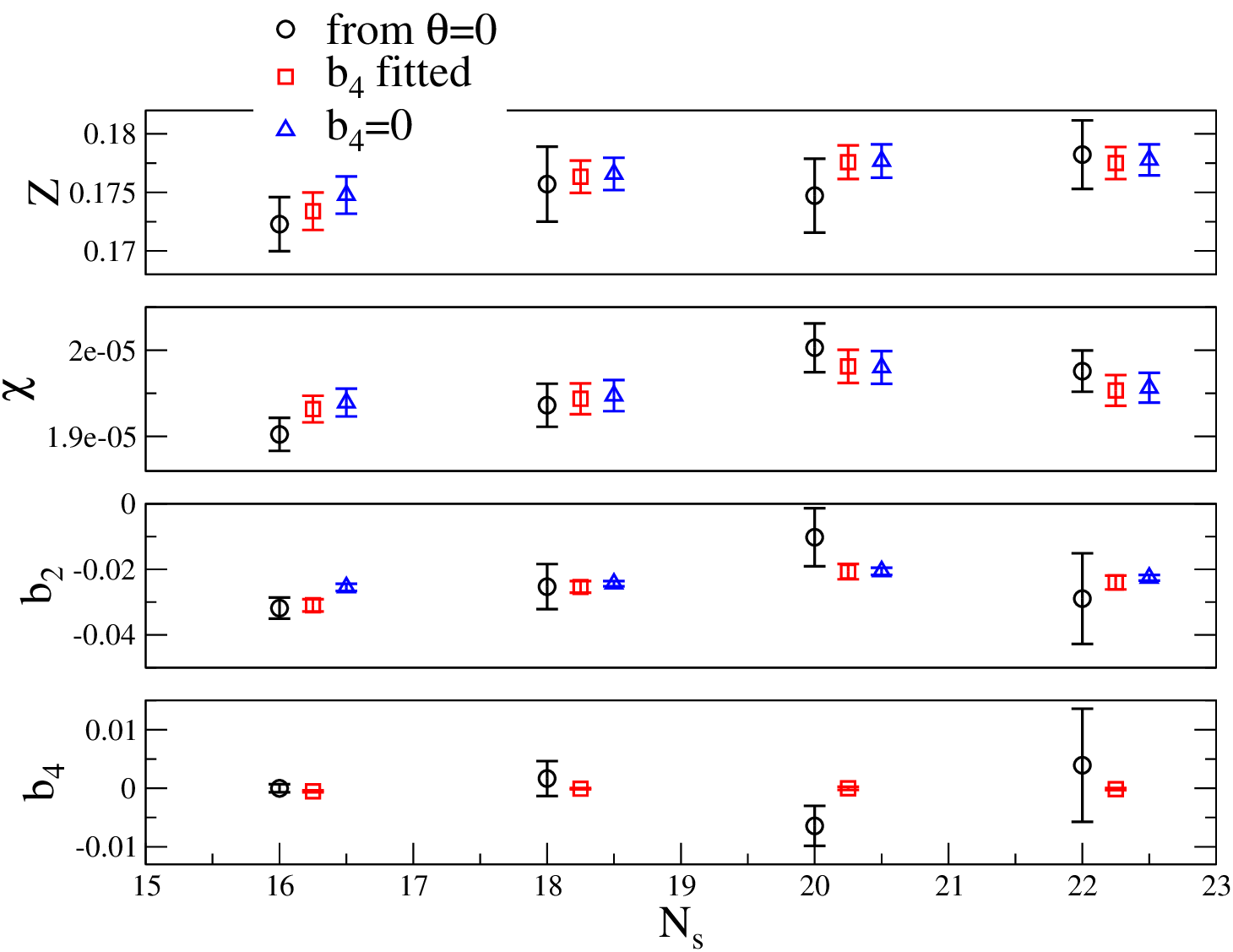}
\caption{Dependence of $Z, \chi, b_2$ and $b_4$ on the lattice size for
$\beta=6.2$. For each lattice size and each observable three estimates are reported:
the one coming just from the $\theta=0$ simulations (i.e. the traditional Taylor expansion
determination), the one obtained by fitting the $b_4$ value and the one obtained by
fixing $b_4=0$.}\label{fig:finiteV}
\end{figure}

In order to investigate finite size effects, we have explored different
lattice sizes $N_s$ for one case, in particular $\beta = 6.2$ corresponding
to a lattice spacing $a \sim 0.068$ fm. In Fig.~\ref{fig:finiteV}
we report the corresponding results obtained both from analytic 
continuation and from a direct measurement of the cumulants at 
$\theta = 0$. We notice that errors on $b_2$ and 
$b_4$ obtained from analytic
continuation scale much better as the volume is increased 
(the underlying reason is discussed in Sec.~\ref{comparison}),
so that in this case one can investigate finite size effects
with much more reliability than with a traditional approach based 
on $\theta = 0$ simulations only. One could try to estimate
the infinite volume limit by explicitly fitting 
the volume dependence of the results, which is expected
to decay exponentially with $N_s$, however 
it is clearly visible 
from the figure that finite size effects are negligible within 
errors for lattices with $N_s \geq 18$, i.e. for
$N_s a > 1.2$ fm. Lattices explored at the other values
of the lattice spacing have been chosen accordingly, i.e.
they correspond to linear sizes
which are well above this threshold ($1.5\,\mathrm{fm}$ at least).

\begin{table}
\begin{tabular}{|l|l|l|l|}
\hline\rule{0mm}{3mm}$\beta$ & $N_t\times N_s^3$ & \# measures & $\tau(Q)$ \\ \hline
\rule{0mm}{3mm}5.96    & $16^4$            & 240K, 130K &  3.47(8) \\ \hline
\rule{0mm}{3mm}6.07    & $22^4$            & 110K, 70K  &  4.8(2)  \\ \hline
\rule{0mm}{3mm}6.20    & $16^4$            & 170K, 60K  &  18(1)   \\ \hline
\rule{0mm}{3mm}6.20    & $18^4$            & 100K, 70K  &  17(1)   \\ \hline
\rule{0mm}{3mm}6.20    & $20^4$            & 100K, 90K  &  17(1)   \\ \hline
\rule{0mm}{3mm}6.20    & $22^4$            & 110K, 80K  &  16(1)   \\ \hline
\rule{0mm}{3mm}6.40    & $30^4$            & 200K, 120K & 214(30)  \\ \hline\hline
\rule{0mm}{3mm}6.173   & $10\times 40^3$   & 50K, 20K   &  27(3)   \\ \hline
\end{tabular}
\caption{Statistics used for the various lattices: the first number of the
third column is the statistics for the $\theta=0$ run, the second is the
typical statistics of the $\theta\neq 0$ runs. The estimated autocorrelation
time of the topological charge is reported in the last column.} 
\end{table}

\begin{table}
\begin{tabular}{|l|l|l|l|}
\hline \multicolumn{4}{|c|}{rounded $Q_L$}\\ \hline
\rule{0mm}{3mm}$\beta$ & $Z$           & $a^4\chi$                  & $b_2$ \\ \hline
\rule{0mm}{3mm}5.95    & $0.12398(31)$ & $1.0744(29)\times 10^{-4}$ & $-0.02457(84)$\\ \hline
\rule{0mm}{3mm}6.07    & $0.15062(62)$ & $4.601(22)\times 10^{-5}$  & $-0.02285(90)$\\ \hline
\rule{0mm}{3mm}6.20    & $0.1778(13)$  & $1.956(17)\times 10^{-5}$  & $-0.02258(86)$\\ \hline
\rule{0mm}{3mm}6.40    & $0.2083(29)$  & $5.94(10)\times 10^{-6}$   & $-0.02347(98)$\\ \hline
\hline \multicolumn{4}{|c|}{non-rounded $Q_L$}\\ \hline
\rule{0mm}{3mm}$\beta$ & $Z$           & $a^4\chi$                  & $b_2$ \\ \hline
\rule{0mm}{3mm}5.95    & 0.13838(36)   & $8.711(23)\times 10^{-5}$  & $-0.01761(68)$ \\ \hline
\rule{0mm}{3mm}6.07    & 0.16300(73)   & $3.940(19)\times 10^{-5}$  & $-0.01898(73)$ \\ \hline
\rule{0mm}{3mm}6.20    & 0.1900(14)    & $1.728(15)\times 10^{-5}$  & $-0.01887(71)$ \\ \hline
\rule{0mm}{3mm}6.40    & 0.2185(28)    & $5.449(88)\times 10^{-6}$  & $-0.02069(89)$ \\ \hline
\end{tabular}
\caption{Results of the combined fit to $Z, \chi$ and $b_2$ using up to the
fourth cumulant of the topological charge.  A vanishing $b_4$ was assumed in the fit.}
\end{table}

\begin{figure}[t!]
\includegraphics[width=0.92\columnwidth, clip]{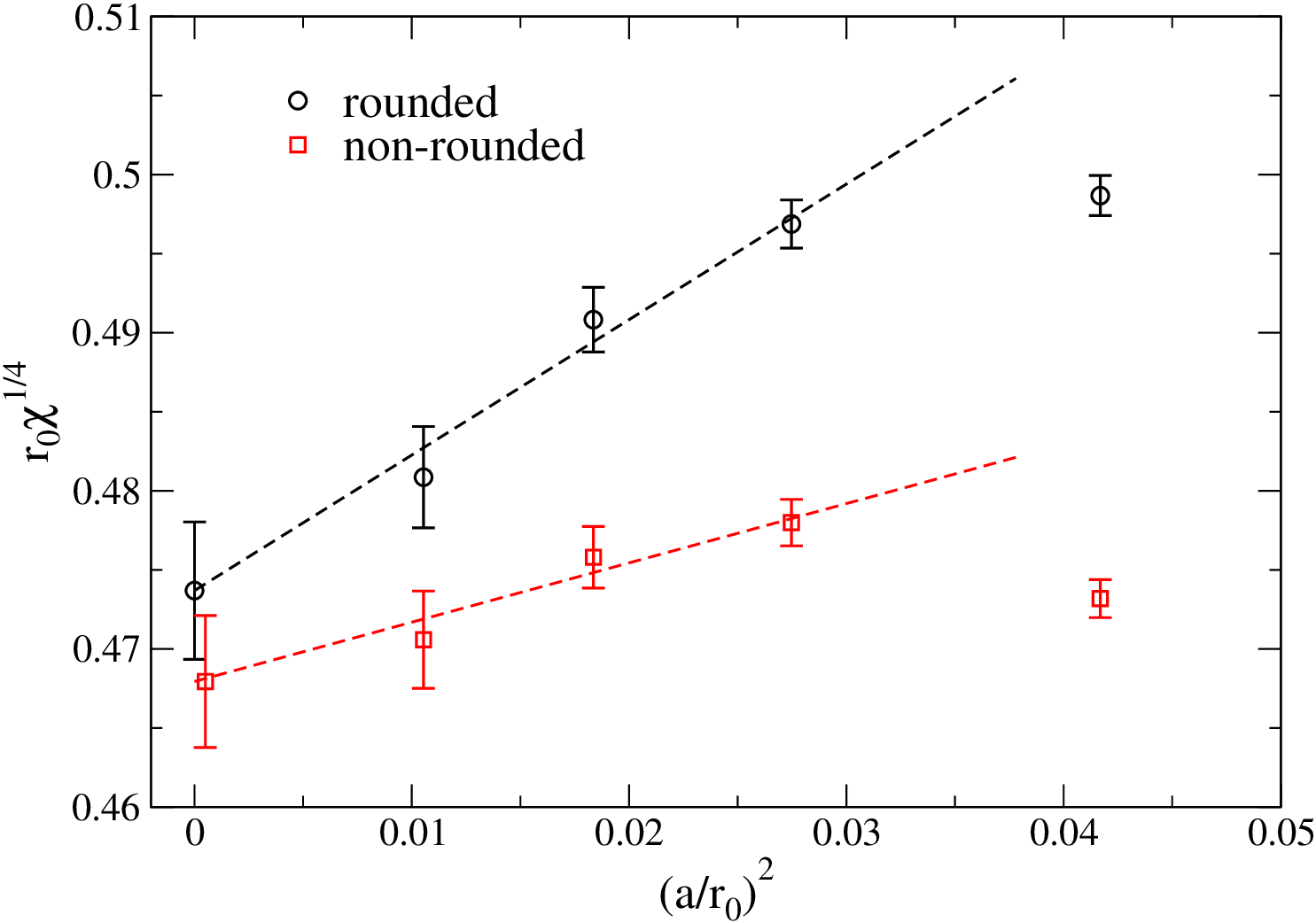}
\caption{Continuum limit for $\chi$,
with and without adopting a rounding to the closest integer
for the topological charge measured after cooling.}\label{fig:chi}
\end{figure}

\begin{figure}[t!]
\includegraphics[width=0.92\columnwidth, clip]{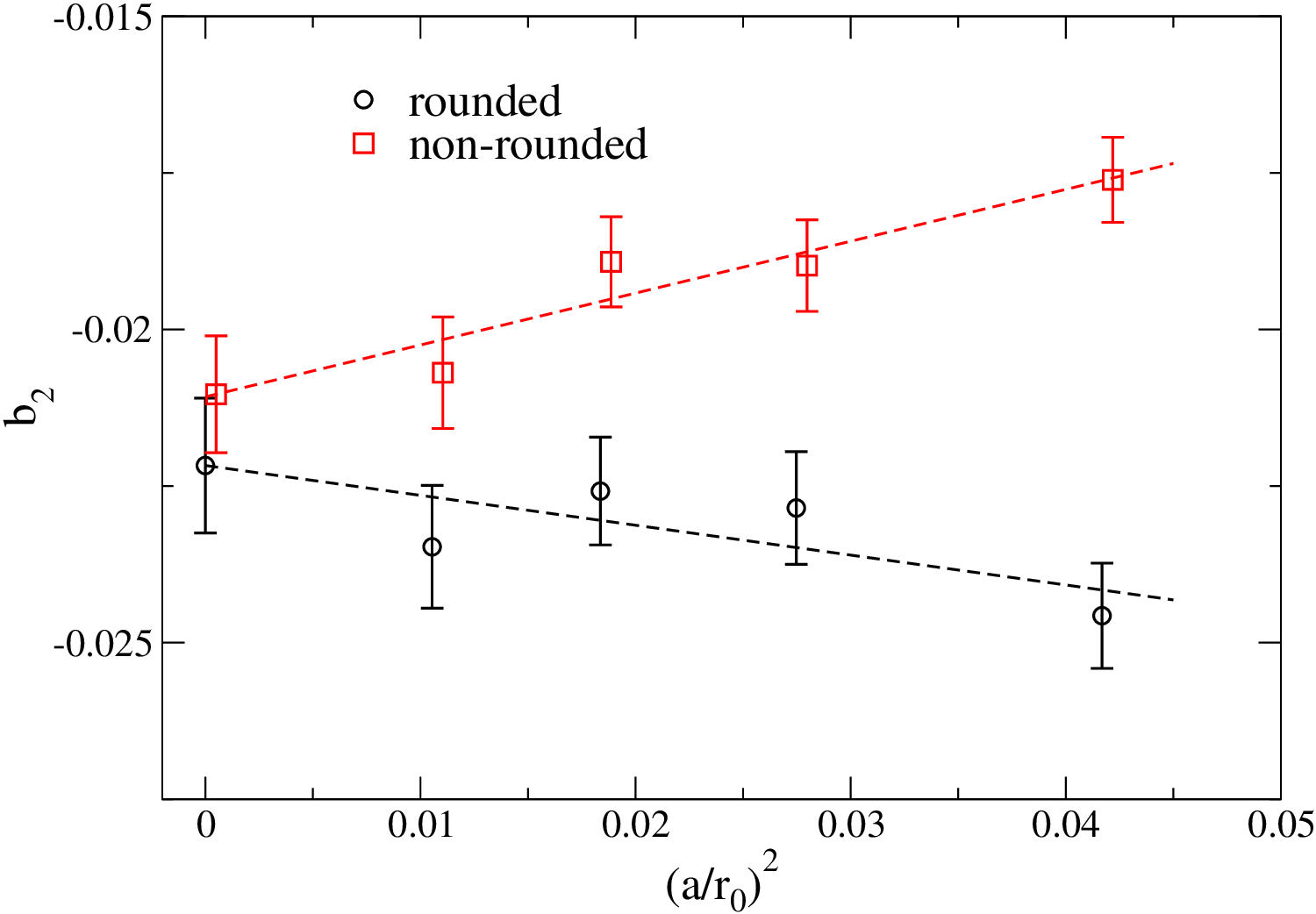}
\caption{Continuum limit for $b_2$,
with and without adopting a rounding to the closest integer
for the topological charge measured after cooling.}\label{fig:b2}
\end{figure}

Let us now go to a discussion of the continuum limit. In order to have better
control on the systematics of the continuum extrapolation, we used two
different definitions of the topological charge, the rounded and the
non-rounded one, see Sec.~\ref{sec:top_lat}.  Since the two definitions are
expected to coincide in the continuum limit and they were
measured on the same set of gauge configurations, 
the differences obtained for the continuum extrapolated values
in the two cases can be used as an estimate of the residual 
systematic uncertainties on the final results.
In Fig.~\ref{fig:chi} we report results obtained for the topological 
susceptibility as a function of $(a/r_0)^2$, where $r_0$ is the 
Sommer scale parameter~\cite{Sommer:1993ce}. In both 
cases (rounded and non-rounded) finite cut-off effects
can be reasonably fitted by a quadratic function of $a$ if
the three finest lattices are considered. Our final estimate is
\begin{equation}
r_0\chi^{1/4}=0.4708(42)(58)
\end{equation}
where the first error is the statistical one and
the second the systematic one;  this value is compatible with previous
determinations \cite{Alles:1996nm, DelDebbio:2002xa, DelDebbio:2004ns,
Durr:2006ky, Luscher:2010ik, Cichy:2015jra, Ce:2015qha} and has a similar
error. The conversion to physical units can be performed e.g.  using
$r_0F_K=0.293(7)$ (from Ref.~\cite{Garden:1999fg}) and the experimental value
$F_k\sim 110\,\mathrm{MeV}$, thus obtaining $\chi^{1/4}=176.8(2.7)(4.2)$. The
first error originates from the uncertainty in $r_0\chi^{1/4}$ (the systematic
and statistical components were summed in quadrature), while the second error
is related to the scale setting and it is the dominant one.

\begin{figure}[tb!]
\includegraphics[width=0.92\columnwidth, clip]{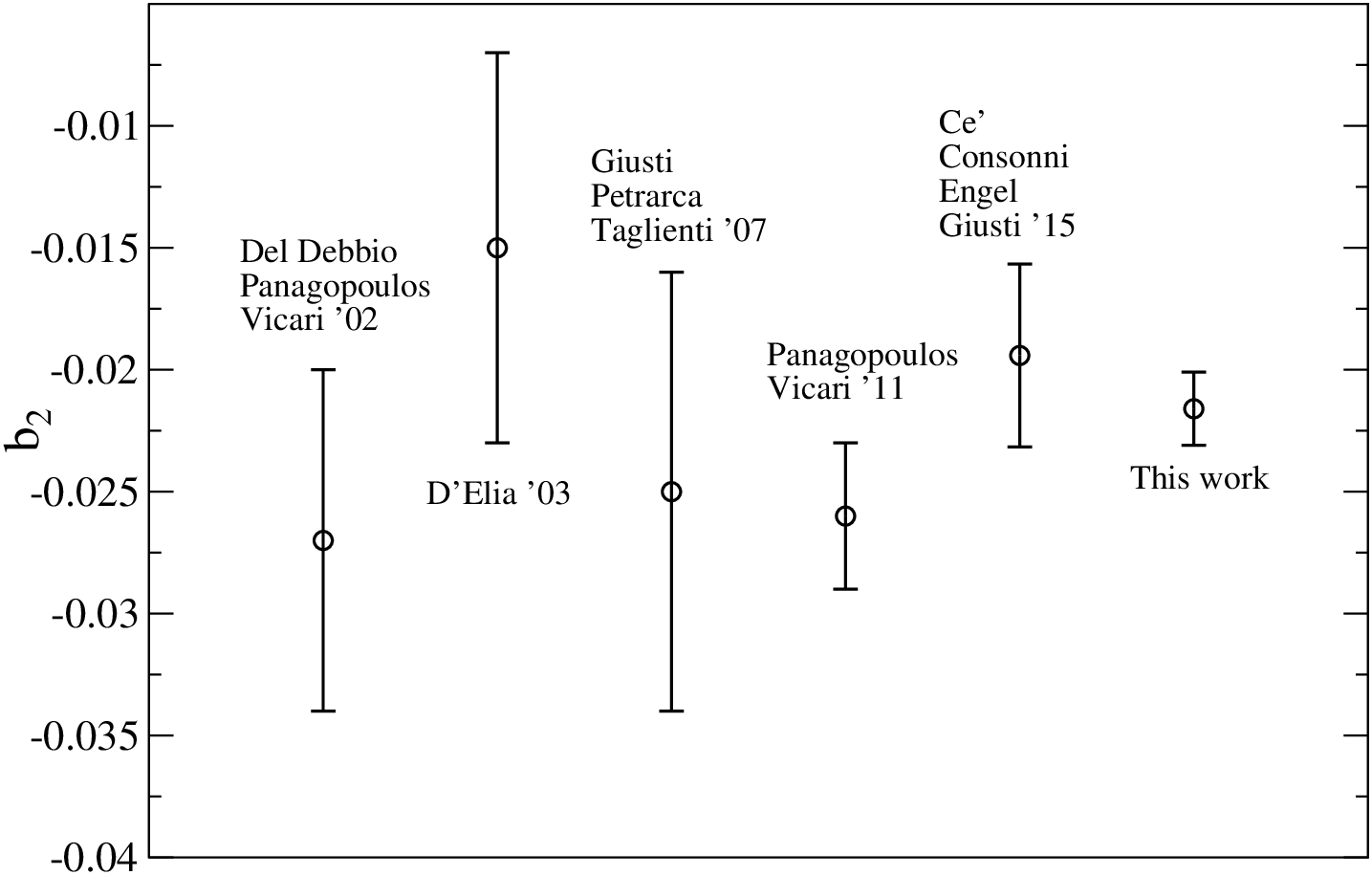}
\caption{Comparison of various determinations of $b_2$ present in the 
literature. From left to right, \cite{DelDebbio:2002xa}, \cite{D'Elia:2003gr},
\cite{Giusti:2007tu}, \cite{pavim}, \cite{Ce:2015qha} and this work.
The error reported for the present determination is a combination
(in quadrature) of the statistical and systematic uncertainties. Notice 
that in Refs.~\cite{Giusti:2007tu} and \cite{Ce:2015qha} results are given
in terms of $R = - 12 b_2$ and have been converted to match with the definition
of the expansion coefficients used in this study.}\label{fig:b2_histo} 
\end{figure}

A similar analysis for $b_2$ is reported in Fig.~\ref{fig:b2}. In this case our
final result is
\begin{equation}
b_2=-0.0216(10)(11)\ ,   
\end{equation}
where again the numbers in parentheses are, from left to right, the statistical
and the systematic error.  This number is compatible with previous results in
the literature~\cite{DelDebbio:2002xa, D'Elia:2003gr, Giusti:2007tu, pavim,
Ce:2015qha}, as can be appreciated from Fig.~\ref{fig:b2_histo} where we report
a summary of all existing determinations, however a sizable error reduction has
been achieved in the present study.

\begin{figure}[tb!]
\includegraphics[width=0.92\columnwidth, clip]{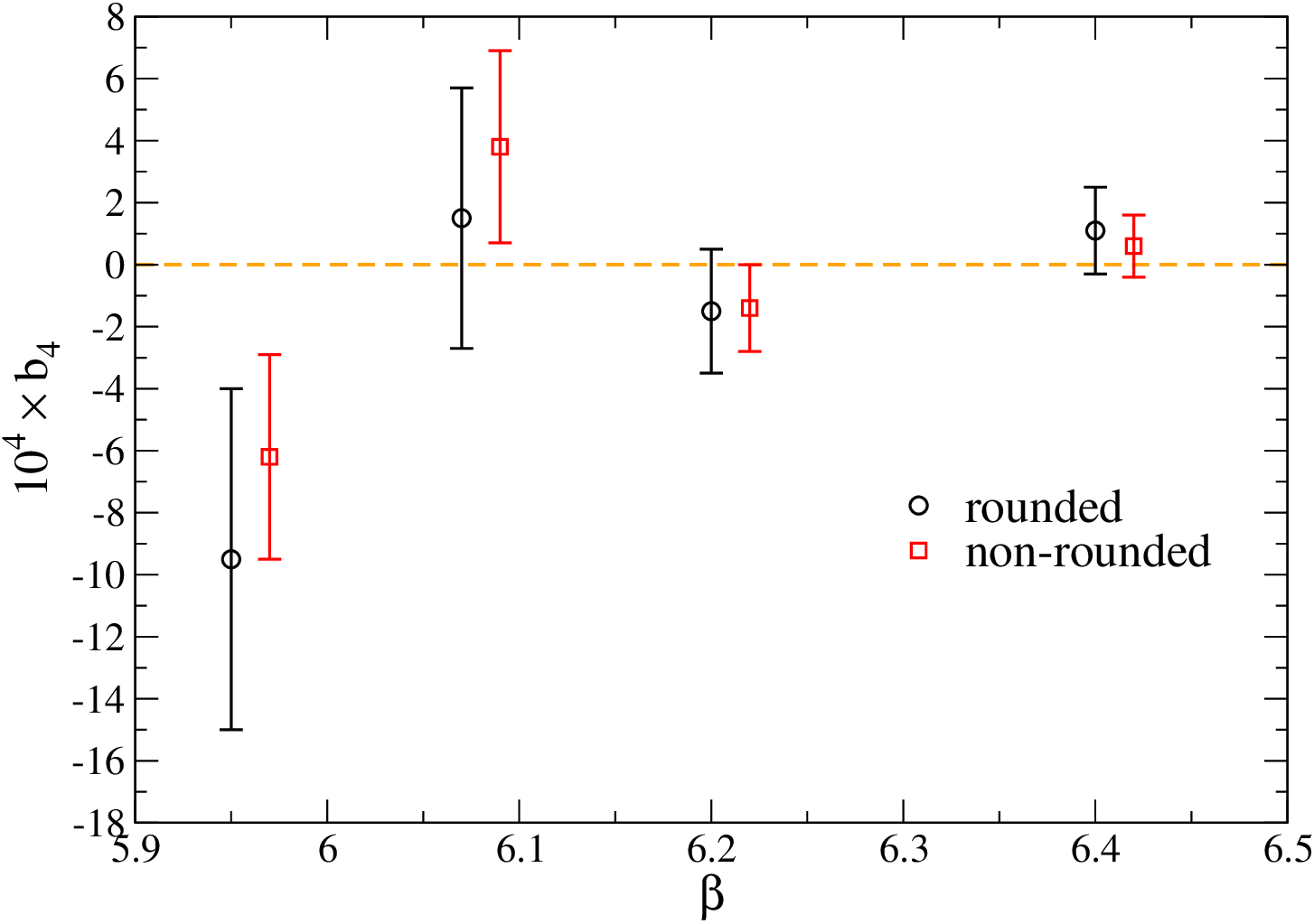}
\caption{Estimates of $b_4$ at various bare couplings.}\label{fig:b4}
\end{figure}

Finally, as we have already emphasized, 
we do not find any evidence of a nonvanishing $b_4$ coefficient, so the best
we can do is to set an upper bound to its absolute value. In Fig.~\ref{fig:b4} we show
the results obtained by using a combined fit for the first four 
cumulants, with
$Z$, $\chi$, $b_2$ and $b_4$ as fit parameters. All data are compatible with zero 
and a reasonable upper bound appears to be 
\begin{equation}
|b_4|\lesssim 4\times 10^{-4}\ .
\end{equation}

\subsection{Comparison with the efficiency of other approaches}
\label{comparison}

As we have already emphasized, Fig.~\ref{fig:finiteV} shows
that, at least for the higher order cumulants, the gain in statistical
accuracy obtained by analytic continuation with respect to a direct
determination at $\theta = 0$ becomes overwhelming as the lattice
volume grows. We are now going to understand why.

The cumulants of the topological charge distribution, 
$\langle Q^{k}\rangle_c$, 
are proportional to the coefficients of the Taylor
expansion of the free energy as a function of $\theta$.
Since the free energy is an extensive quantity, the cumulants
are extensive as well, therefore 
$\langle Q^{k}\rangle_c$ scales proportionally to the
four-dimensional volume ${\cal V}$, and the ratios of cumulants,
i.e. the coefficients $b_2, b_4, \dots$ entering the free energy
density, are volume independent.
On the other hand, in the 
large volume limit the probability distribution
of the topological charge is dominated by Gaussian fluctuations
with variance $\chi {\mathcal V}$, hence the typical fluctuation
has size $\delta Q\sim \sqrt{\chi \mathcal{V}}$. 
Since at $\theta = 0$ the distribution of $Q$ is centered around zero,
one has that the terms that sum up to $\langle Q^{k}\rangle_c$ (i.e.
$\langle Q^{k}\rangle, \langle Q^{2}\rangle\langle Q^{k-2}\rangle, \ldots
\langle Q^2\rangle^{k/2}$) grow like $(\chi \mathcal{V})^{k/2}$.
Therefore the cumulant itself, which is of order $\chi \mathcal{V}$, comes out
from the precise cancellation of higher order terms, so that the 
signal-to-noise ratio is expected to scale like
\begin{equation*}
(\chi \mathcal{V})/(\chi \mathcal{V})^{k/2} = (\chi \mathcal{V})^{(2-k)/2}\, .
\end{equation*}
This is consistent with the fact that the cumulants measure the deviations
from a purely Gaussian distribution, and such deviations, 
because of the central
limit theorem, 
become less and less visible as the infinite volume limit is approached.

We conclude that the error expected in the standard 
determination of $b_{2n}$ through the cumulants at $\theta = 0$ is of the 
order of $(\chi \mathcal{V})^{n}/\sqrt{N}$, where
$N$ is the size of the statistical sample, i.e. the number of independent
gauge configurations. 
This is a particularly severe case of lacking of self-averaging~\cite{MBH},
for instance in the case of $b_2$ one has to increase the number
of measurement proportionally to $\mathcal{V}^2$, 
in order to keep a fixed statistical accuracy as the volume is increased.
Notice that this problem is not marginal. Indeed, 
the important parameter entering this scaling is the 
combination $\chi \mathcal{V}$:
since $\chi^{1/4} \sim 180\, \mathrm{MeV} \sim (1.1\,\mathrm{fm})^{-1}$, we expect 
a strong degradation in the signal-to-noise ratio as one approaches 
lattices which exceed one fermi in size, which is consistent 
with our observations; unfortunately, this is also
at a threshold where finite size effects are still important.

On the contrary, when using the analytic continuation approach,
this problem disappears. Indeed, in this case the information 
about each free energy coefficient is contained also in the
$\theta$ dependence of the lowest order cumulants, including
$\langle Q\rangle_{c,\theta_I}$, whose signal to noise ratio
scales with volume as $\mathcal{V}^{1/2}$.  
The improvement is related to the fact that one is probing 
the system with an explicit non-zero external source, and is similar to 
that obtained when
switching from the measurement of 
fluctuations to the measurement of the magnetization as a function 
of the external field in the determinations of the magnetic susceptibility
of a material.

Therefore, one expects that the final statistical accuracy at fixed number
of measurements, improves when increasing the volume, rather than
degrading, like for standard self-averaging quantities.
Actually, in the particular case of the
analytic continuation in $\theta$, there is a slight complication related to
 the renormalization constant $Z$. 
This can be computed as in Eq.~\eqref{eq:Zdet}, but in
this case we need also $\langle Q^2\rangle$, whose precision scales like
$\mathcal{V}^{0}$ with volume.  The method of global fit is not qualitatively
better in this respect: since $Z$ appears as a rescaling of $\theta$ it cannot
be determined just by using the first cumulant, its value is fixed by the
comparison of at least $\langle Q\rangle_{c,\theta_I}$ and $\langle
Q^2\rangle_{c,\theta_I}$ and the precision of the second scales as
$\mathcal{V}^0$. We thus see that the best we can obtain with the analytic
continuation approach is a scaling of the precision of $b_{2n}$
with volume like $\mathcal{V}^0$; this is suboptimal with respect to the naive
expectation $\mathcal{V}^{1/2}$, 
but still a tremendous improvement with respect
to the Taylor expansion method. On the other hand, the precisions of $Z$ and
$\chi$ scale in both approaches as $\mathcal{V}^0$.

All these scalings, and the improvement gained by analytic continuation, can be
nicely seen in Fig.~\ref{fig:finiteV}. Three determinations of $Z$, $\chi$ and
$b_2$ are shown: the first one is obtained by using the Taylor expansion
method, i.e using just the $\theta=0$ sets, the second one uses the analytic
continuation approach, including in the global fit up to the fourth cumulant of
the topological charge and fitting $Z$, $\chi$, $b_2$ and $b_4$, the last one
follows the same strategy as the second one but $b_4$ is fixed to zero in the
fit. The determination of $b_4$ obviously used only the first two methods.
Actually, a reasonable comparison between different methods should proceed
along the following lines: compare the errors obtained by different approaches
when using the same machine-time. For the comparison to be fair we thus have to
rescale the statistics of the $\theta=0$ determination by about a factor of
$10$, i.e. the errors of the corresponding determination by about a factor of
$3$. We thus see that no gain is obtained by using the analytic continuation
approach to determine $Z$ and $\chi$, while it is extremely beneficial for
$b_2$ and $b_4$, as theoretically expected.  \\

\begin{figure}[tb!]
\includegraphics[width=0.92\columnwidth, clip]{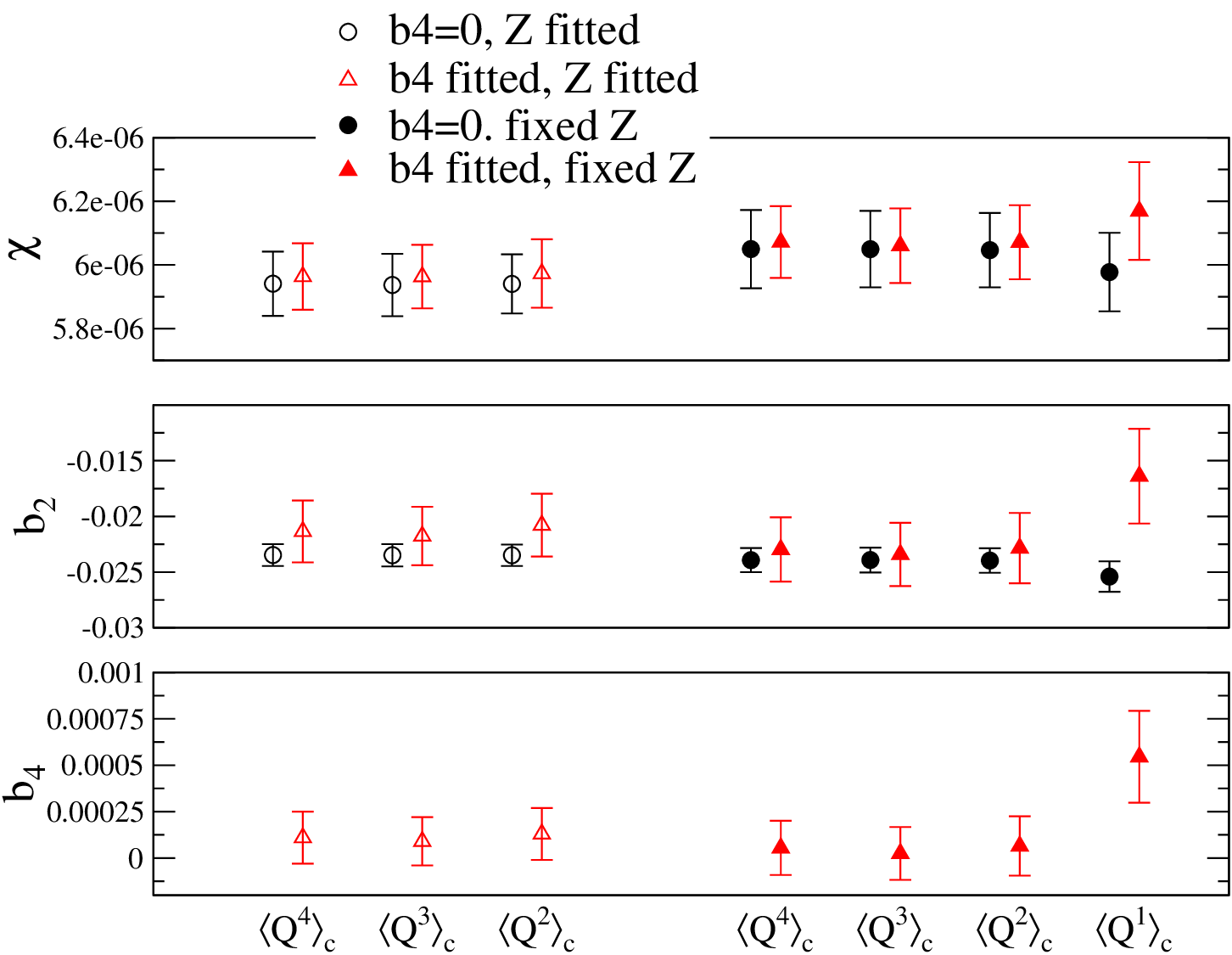}
\caption{Comparison between different fitting procedures in the analytic continuation
approach.  Data refer to the $30^4$ lattice at bare coupling $\beta=6.40$.
Empty symbols are the result of a global fit procedure (in which also $Z$ is
determined) using up to the $4^{th}$, $3^{rd}$ or $2^{nd}$ cumulant. Full symbols are
obtained using a global fit (with $Z$ fixed by Eq.~\eqref{eq:Zdet}) and using up to
the $4^{th}$, $3^{rd}$, $2^{nd}$ or $1^{st}$ cumulant.}\label{fig:comparefit}
\end{figure}

A last point that remains to be investigated is the effectiveness, within the
analytic continuation approach, of the global fit to all the cumulants with
respect to the traditional procedure of using just $\langle Q\rangle_c$, with
$Z$ fixed using Eq.~\eqref{eq:Zdet}. Such a comparison is performed in
Fig.~\ref{fig:comparefit}, in which also intermediate cases are shown: a global
fit of all the parameters using different numbers of cumulants or a global fit
of all the parameters but $Z$ (which was fixed by Eq.~\eqref{eq:Zdet}), using
different numbers of cumulants. As can be seen the inclusion in the fit of
cumulants of order higher than the second does not improve appreciably the precision
of the result; the error reduction attainable by fitting also $Z$ instead of
fixing it before the fit is not huge, but still significant. Indeed, by using all the
cumulants of the topological charge and fitting $Z$ we reduced the errors by
about a factor of $2$ with respect to the traditional procedure of using just
$\langle Q\rangle_{\theta_I}$ with $Z$ fixed by Eq.~\eqref{eq:Zdet}.

\subsection{Some considerations about the finite temperature case}
\label{finiteT}

A strategy allowing to measure $b_{2n}$ on arbitrary 
large volumes represents a substantial improvement, that
permits to gain in statistical accuracy and 
reduces the possible systematics related to finite size effects.
 There are however cases in which this possibility is not just an
``improvement'', but it enables the study of otherwise intractable problems.
This is the case of $b_2$ at finite temperature for $T<T_c$. Once the lattice
spacing is fixed and the temperature is fixed (i.e. $\beta$ and $N_t$ are
fixed), to ensure that we are studying a finite temperature system we have to
impose the constraint $N_s\gg N_t$ (typically $N_s\gtrsim 4N_t$ is used) and
this fixes a lower value for the acceptable four-volumes to be used.  What
typically happens is that this minimum value of $\mathcal{V}$ is large enough
to make a measure of $b_2$ using the Taylor expansion method extremely
difficult.

\begin{figure}[tb!]
\includegraphics[width=0.92\columnwidth, clip]{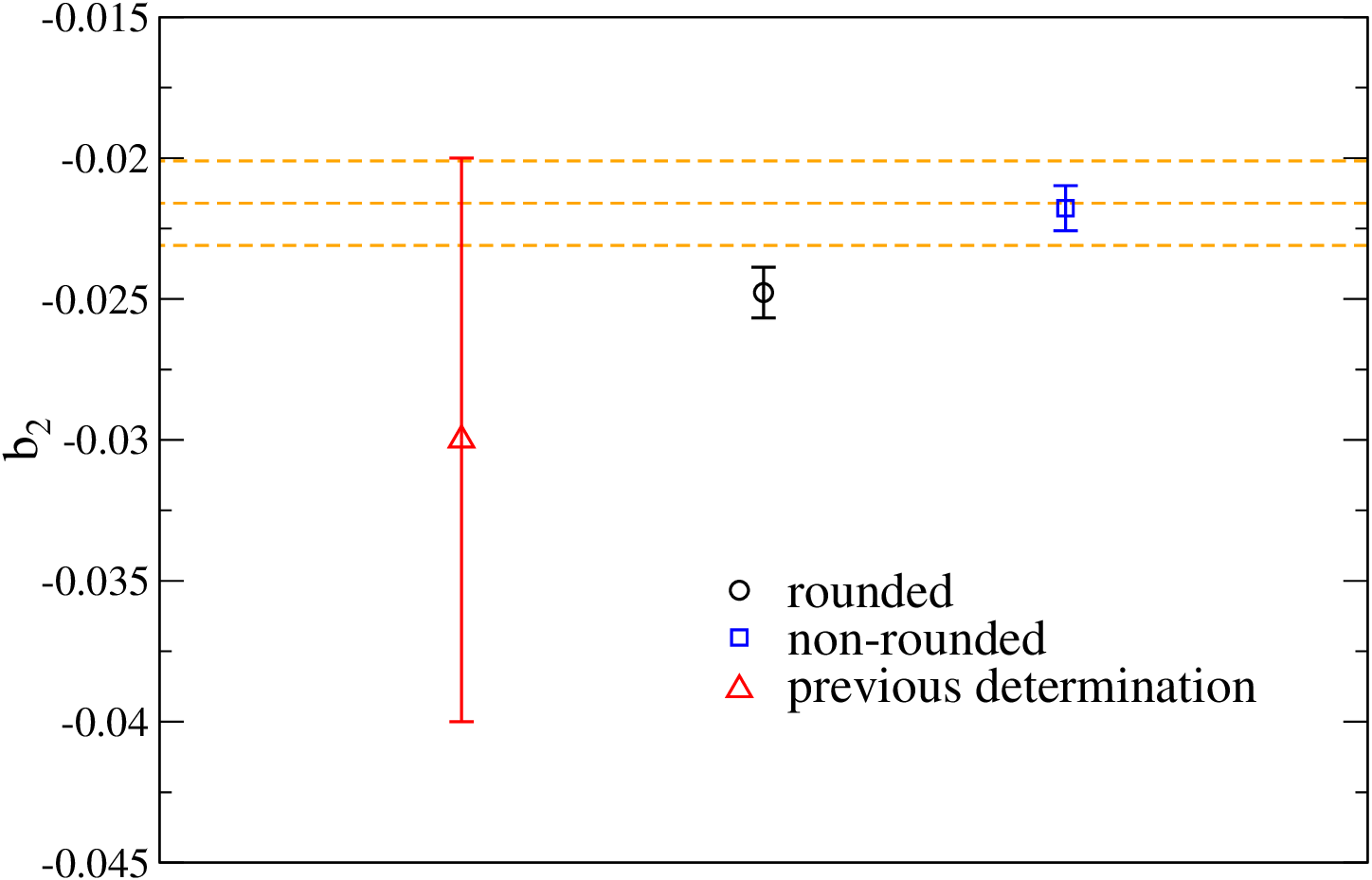}
\caption{$b_2$ value at $T\simeq 0.95T_c$ on a $10\times 40^3$ lattice (bare
coupling $\beta=6.173$), with a comparison of the determination in
\cite{Bonati:2013tt}, obtained using the Taylor expansion method, and a new
estimate obtained by using the analytic continuation method.
The horizontal lines denote the $T=0$ result.}\label{fig:finiteT}
\end{figure}

In order to verify that the analytic continuation approach works equally well
also at finite temperature, we performed a simulation at $\beta=6.173$ on a
lattice $10\times 40^4$. Such a point was previously investigated in
\cite{Bonati:2013tt} using the Taylor expansion method.
The result from Ref.~\cite{Bonati:2013tt}, 
together with our new determinations
and the $T = 0$ result are reported in
Fig.~\ref{fig:finiteT}.
The old determination of Ref.~\cite{Bonati:2013tt} was based 
on a statistics of about $800K$ measures: our present determination
obtained via analytic continuation reaches 
a precision which is more than one order of magnitude larger, with about 
half the computational effort (in particular, a machine-time 
equivalent to around $350K$ measures taken at $\theta=0$).

Different considerations should be made for temperatures above 
the critical temperature $T_c$. Indeed, 
in all the recent studies of $b_2$ at finite temperature \cite{Bonati:2013tt,
Bonati:2015uga, Borsanyi:2015cka, Xiong:2015dya} (in which the Taylor expansion
method was always used) it was noted that the determination of $b_2$ in the low
temperature phase is very difficult, while it gets abruptly easier above
deconfinement. The reason for this fact is simple: we have
seen that the relevant parameter for the degradation of the statistical
accuracy in the determination of the cumulants at $\theta = 0$ 
is $\chi \mathcal{V}$. Since at
the deconfinement transition the topological susceptibility steeply 
decreases~\cite{Alles:1996nm, Gattringer:2002mr, Lucini:2004yh, DelDebbio:2004rw,
Berkowitz:2015aua, Kitano:2015fla, Borsanyi:2015cka, Xiong:2015dya}, a precise
determination of $b_2$ by the standard Taylor expansion approach 
becomes much easier, in the
deconfined phase, even on moderately large volumes.

\section{Conclusions}

In this study we have exploited numerical simulations performed
at imaginary values of $\theta$ and analytic continuation in order 
to determine the dependence of the free energy density of the
$SU(3)$ pure gauge theory on the topological parameter $\theta$.
As an improvement with respect to previous applications of the same
strategy~\cite{pavim}, we have considered a global fit 
to the $\theta$ dependence of various free energy derivatives, i.e.~various 
cumulants of the topological charge distribution.
That has permitted us to reach an increased precision, obtaining 
the following estimates for the continuum extrapolated values:
$r_0\chi^{1/4}=0.4708(42)(58)$, $b_2=-0.0216(10)(11)$ 
and $|b_4|\lesssim 4\times 10^{-4}$.

The strategy based on analytic continuation turns out to be particularly 
well suited for the determination of the higher order terms,
for which the traditional Taylor expansion method, based on the determination
of cumulants of the topological charge distribution
at $\theta = 0$, faces a severe problem when approaching large volumes,
where corrections to Gaussian-like fluctuations become hardly measurable.
Indeed, we have shown that the statistical error attained
in the determination of 
$b_{2n}$ through the cumulants at $\theta = 0$ scales with 
the four-dimensional volume $\mathcal{V}$, for a fixed number of measurements,
like $(\chi \mathcal{V})^{n}$, where
$\chi$ is the topological susceptibility. As we have 
shown, this property
of analytic continuation turns out to be essential for the determination 
of $\theta$-dependence right below the deconfinement temperature
$T_c$, hence this strategy could be adopted for a 
future improved determination of the change of 
$\theta$-dependence taking place at deconfinement~\cite{Bonati:2013tt}.

\acknowledgments
We thank Francesco Negro and Ettore Vicari for useful discussions.
Numerical simulations have been performed on the CSN4 cluster of the Scientific
Computing Center at INFN-PISA and on the Galileo machine at CINECA (under INFN
project NPQCD).  Work partially supported by the INFN SUMA Project.

\end{document}